\documentclass[amsmath,amssymb,prl,nofootinbib]{revtex4}

\usepackage[dvips]{color}
\usepackage{epsfig}
\usepackage[latin1]{inputenc}

\def\be{\begin{equation}}
\def\ee{\end{equation}}
\def\ba{\begin{array}}
\def\ea{\end{array}}
\def\bea{\begin{eqnarray}}
\def\eea{\end{eqnarray}}

\newcommand{\nc}{\newcommand}
\nc{\tcb}{\textcolor{blue}}
\nc{\tcr}{\textcolor{red}}
\nc{\tcg}{\textcolor{green}}

\nc{\qq}{\qquad\qquad}
\nc{\dis}{\displaystyle}
\nc{\ug}{\; = \;}
\nc{\Ebf}{\mbox{\boldmath $E$}}
\nc{\Bbf}{\mbox{\boldmath $B$}}
\nc{\abf}{\mbox{\boldmath $a$}}
\nc{\vbf}{\mbox{\boldmath $v$}}
\nc{\Fbf}{\mbox{\boldmath $F$}}
\nc{\rbf}{\mbox{\boldmath $r$}}
\nc{\Jbf}{\mbox{\boldmath $J$}}
\nc{\rd}{{\rm d}}
\nc{\dtau}{{\rd\tau}}
\nc{\dt}{{\rd t}}
\nc{\omf}{{\frac{\omega}{\omz}}}
\nc{\tz}{\tau_0}

\begin{document}


\vspace{1truecm}

\title[Fundamental times, lengths and physical constants]{Fundamental times, lengths and physical constants:\\ some unknown contributions by Ettore Majorana}

\author{S. Esposito}
\address{Dipartimento di Scienze Fisiche,
Universit\`a di Napoli ``Federico II'' and I.N.F.N. Sezione di
Napoli, Complesso Universitario di Monte S. Angelo, Via Cinthia,
80126 Naples, Italy}
\email{Salvatore.Esposito@na.infn.it}

\author{G. Salesi}
\address{\mbox{Facolt\`a di Ingegneria, Universit\`a di Bergamo, viale Marconi 5, 24044 Dalmine (BG) Italy}\\ 
and I.N.F.N. Sezione di Milano, via Celoria 16, 20133 Milan, Italy}%
\email{salesi@unibg.it}

\begin{abstract}
We review the introduction in physics of the concepts of an elementary space length and of a fundamental time scale, analyzing some related unknown contributions by Ettore Majorana. In particular, we discuss the quasi-Coulombian scattering in presence of a finite length scale, as well as the introduction of an intrinsic (universal) time delay in the expressions for the retarded electromagnetic potentials. Finally, we also review a special model considered by Majorana in order to deduce the value of the elementary charge, in such a way anticipating key ideas later introduced in quantum electrodynamics.
\end{abstract}

\maketitle



\section{Introduction: The known story}


\noindent The idea of elementary space or time intervals has been quite a recurring one in the philosophical and scientific literature of every time and, as almost all the fundamental physical ideas, from time to time it has been recovered from oblivion. Typical examples range from the ancient Greek period to our days, namely from Zeno's paradox on Achilles and the tortoise to Poincar\'e and Mach, who regarded the concept of continuum as a consequence of our physiological limitations.\footnote{``...le temps et l'espace ne reprèsentent, au point de vue physiologique, qu'un continue apparent, qu'ils se composent très vraisemblablement d'elements discontinus, mais qu'on ne peut distinguer nettement les uns des autres.'' \cite{MachPoin}}

As often shown in the widespread literature on the structure of particles (e.g., see below, the ``chronon theory''),
a time-discretization for an elementary object involves  an ``internal'' (that is, referred to the center-of-mass frame) motion associated with microscopic space distances. For such an elementary particle, an ``extended-like'' structure, rather than a pointlike one, is expected and this has important consequences even for the fundamental theory describing that particles since, indeed, the classical theory of a pointlike charged particle leads to well-known divergency problems, which are only apparently solved by the renormalization group theory.

In the realm of particle physics, the concepts of an elementary time and length are substantially equivalent for several reasons: in most cases one can pass from time to space just by introducing the speed of light. Already in 1930's (see below), it was early realized that a fundamental length is directly related to the existence of a ``cut-off'' in the momentum transferred during the particle-detector interaction, which is required in order to avoid  the emergence of the so-called ultraviolet catastrophe in quantum field theories. Another way to provide a small distance cut-off in field theory is to formulate it on a discrete lattice: this approach was introduced in 1940 by Wentzel \cite{Wentzel}, but only later it went into detail, becoming a standard  technique in modern numerical approach to high energy interactions, as e.g. in quantum chromodynamics.

Space-time discretizations, or fundamental scales, are introduced in classical and quantum theories mainly for two reasons: to prevent undesired infinities in some physical quantities, or (though less known) to explain the emergence of an intrinsic angular momentum for spinning particles.

\

\noindent The first appearance in physics of a fundamental length was suggested by H.A. Lorentz, who introduced the so-called \textit{classical electron radius} $R_{\rm cl}\equiv e^2/mc^2 \simeq 2.82\cdot 10^{-13}$\,cm in his famous classical relativistic model of the electron;\footnote{Lorentz considered electrons just as rigid spheres, albeit contracting when moving in the ether \cite{Lorentz}.} it is equivalently known as ``Lorentz radius'' or ``Thomson scattering length''.
Roughly speaking, such a radius corresponds to the electron size required for obtaining its mass from the electrostatic potential energy, assuming that quantum mechanical effects are irrelevant. Actually,
such effects play a non-negligible role in the behavior of charged particles at very short distances, and the classical electron radius may be considered as the length scale at which renormalization effects become relevant in quantum electrodynamics. In other words, $R_{\rm cl}$ denotes the watershed between classical and quantum electrodynamics; it appears in the semiclassical theory for the non-relativistic Thomson scattering, as well as in the relativistic Klein-Nishina formula \cite{KN}.

\

\noindent In early semiclassical models, the electron was substantially regarded as a charged sphere but, unfortunately, the electromagnetic self-energy of such a sphere diverges in the limit of pointlike charge. This was, then, the basic motivation that led M.\,Abraham, Lorentz and, later, P.A.M.\,Dirac, to derive an equation of motion for the electron where a finite, non-vanishing size is fully taken into account \cite{ALD}. In such a way a fundamental length was effectively introduced in order to properly describe the motion of charged particles. Particularly extensive work on extended, not pointlike, charges was carried out by A.\,Sommerfeld \cite{Sommerfeld}. In studying the motion of a sphere of radius $R$ with a uniform surface charge, he obtained the following ``retarded-like'' first-order differential-finite difference equation ($T\equiv2R/c$):
\be
m\abf \ug e(\Ebf + \vbf\times\Bbf) + \frac{2}{3}\frac{e^2}{Rc^2}\,\frac{\vbf(t-T)-\vbf(t)}{T}\,,
\ee
where the electromagnetic field of the charge itself causes a self-force, which has a delayed effect on its motion. The ``delay time'' $2R/c$ appearing in the Sommerfeld equation, corresponding to the time the light takes to go across the sphere, was one of the earlier elementary times introduced in microphysics. Later in the 1920s, J.J.Thomson \cite{Thomson} suggested that the electric force may act in a discontinuous way, producing finite increments of momentum separated by finite intervals of time.

The relativistic version of Sommerfeld's non-relativistic approach was developed by P.\,Caldirola \cite{CALREV4}. His theory of the electron is one of the first and simplest theories which assumes {\it a priori} a minimum time interval: it is based on the existence of an elementary proper-time interval, the so-called \textsl{chronon}. When applied to electrons in an external electromagnetic field, such a finite-difference theory succeeds (already at the classical level) in overcoming all the known difficulties met by the Abraham-Lorentz-Dirac (ALD) theory, like the so-called pre-acceleration problem and the emergence of run-away solutions \cite{Elieser,Ashauer}. It is also able to give a clear answer to the problem related to some ambiguities associated with the hyperbolic motion of the electron \cite{LFR}, as well as to the question of whether a free falling charged particle does or does not emit radiation \cite{Rohr}. Furthermore, at the quantum level, Caldirola's chronon theory is seemingly able to explain the origin of the ``classical (Schwinger's) part'', $e\hbar/2mc\cdot\alpha/2\pi = e^3/4\pi mc^2$, of the anomalous magnetic momentum of the electron as well as the mass spectrum of the charged leptons \cite{CALREV4}.
In this theory, time flows continuously, but, when an external force acts on the electron, the reaction of the particle to the applied force is not continuous: the value of the electron velocity
$u_\mu$ is supposed to jump from $u_\mu(\tau - \tau_0)$ to $u_\mu(\tau)$ only at certain positions $s_{n}$ along its world line. These ``discrete positions'' are such that the electron takes a time $\tau_0$ for traveling from one position $s_{\rm{n} - 1}$ to the next one $s_{n}$. In principle, the electron is still considered as pointlike, but the ALD equation for the relativistic radiating electron is replaced: \ (i) by a corresponding {\em finite--difference} (retarded) equation in the velocity $u^\mu(\tau)$
\begin{eqnarray}
& & {{m} \over {\tau_0}}\left\{ {u^\mu \left( \tau  \right)-u^\mu \left( {\tau -\tau_0} \right)+{{u^\mu \left( \tau  \right)
u^\nu \left( \tau  \right)} \over {c^2}}\left[ {u_\nu \left( \tau
\right)-u_\nu \left( {\tau -\tau_0} \right)} \right]} \right\} \nonumber \\
& & = {e \over c}F^{\mu \nu}\left( \tau  \right)u_\nu \left( \tau  \right)
\end{eqnarray}
(which reduces to the ALD equation when $\tau_{0}\ll\tau$) and \ (ii) by a second equation (the so-called {\em transmission law}), connecting the discrete positions $x^\mu(\tau)$ along the world line of the particle among themselves:
\be
x_\mu \left( {n\tau_0} \right)-x_\mu \left( (n-1) \tau_0 \right) =
{\displaystyle{\tau_0\over 2}}\left\{ {u_\mu \left( {n\tau_0} \right)-u_\mu
\left[{\left( {n-1}\right)\tau_0} \right]}\right\}\,,
\ee
which is valid inside each discrete interval $\tau_{0}$ and describes the ``internal'' motion of the electron. In such equations the chronon associated with the electron results to be \ $\dis{\tau_0 \over 2} \equiv \theta_0 = {2 \over 3}{{k e^2} \over {m c^3}} \simeq
6.266 \cdot 10^{-24} {\rm s}$ ($k \equiv 1/4\pi\varepsilon_0$),  depending, therefore, on the particle properties (namely, on its charge $e$ and on its rest mass $m$).

\

\noindent A different elementary length appearing in modern physics is the \textit{Compton wavelength}, introduced in 1923 by A.H.\,Compton \cite{Compton} in his explanation of the scattering of photons by electrons. It is defined as $\dis\lambda_{\rm C}\equiv h/mc$, and corresponds to the wavelength of a photon whose energy equals the particle rest mass; for the electron, the Compton wavelength is about $2.42\cdot 10^{-10}$\,cm. The meaning of this length is quite clear: it indicates the watershed between classical mechanics and quantum wave-mechanics, since for spatial distances below $h/mc$ a microsystem behaves as a very quantum object. This was promptly realized by E.\,Schr\"odinger in 1930-31 \cite{Schroedinger} in his study of the so-called \textsl{Zitterbewegung}, which emerges in the Dirac theory by means of the Darwin term appearing in the (decomposed) Hamiltonian of the Dirac equation, as found by C.G.\,Darwin \cite{Darwin} already in 1928. Indeed, because of the Zitterbewegung (see also below) the electron undergoes extremely rapid fluctuations on scales just of the order of the Compton wavelength, causing, for example, the electrons moving inside an atom to experience a smeared nuclear Coulomb potential (this peculiar form of microscopic kinetic energy is presently better known as ``quantum potential''of the Schr\"odinger equation \cite{QP}).

The Compton wavelength plays also the role of the spatial scale which naturally arise in any semi-classical model of spinning particles, when attempting to formulate a kinematical picture of the intrinsic angular momentum. Already in the 1920s, another extended-like structure of particles (in addition to the electrodynamical one mentioned above) was discovered, namely the \textsl{mechanical} (or, rather, kinematical) configuration in the particle's center-of-mass frame due to the spin. Indeed, in 1925 R. Kronig, G. Uhlenbeck and S. Goudsmit \cite{U-G} introduced the celebrated \textsl{self-rotating electron} hypothesis, suggesting a physical interpretation for particles spinning around their own axis. Now, it is interesting to notice that any mechanical model of a self-rotating electron fatally leads to a fundamental length for this very peculiar rotator, resulting to be just the Compton wavelength $\lambda_{\rm C}$. Anyway, models with a self-rotating spinning charge are very naive and problematic.\footnote{It is quite famous a statement by A.O. Barut: ``If a spinning particle is not quite a point particle, nor a solid three dimensional top, what can it be?'' \cite{BB}} In the subsequent literature, these problems were overcome by considering kinematical theories of the spin based on the Zitterbewegung \cite{RECSAL,ZBW}, where spinning particles, though not effectively extended, can be considered as something which is half-way between a point and a rotating body (as, e.g., a top).
In these models the center of the (even pointlike) charge is spatially distinct from the center-of-mass, so that velocity and momentum are no longer parallel vectors, and an internal (microscopic), rotational spin motion of the particle arises in addition to the external (macroscopic) translational motion of the center-of-mass. 
As a consequence, in the classical limit the global inertial motion of a free spinning particle turns out to be no longer uniform and rectilinear, but rather accelerated and helical, with a radius once again equal to the Compton wavelength.

\

\noindent A different insight was provided by quantum mechanics with its discrete energy levels and the uncertainty principle. This led physicists to speculate about space-time discreteness as early as
the 1930's. W. Heisenberg himself \cite{Heisenberg}, for example, in 1938 noted that physics {\it must} have a fundamental length scale which, together with $h$ and $c$, might allow the derivation of the mass of particles. In Heisenberg's view, this length scale would have been around $10^{-13}$cm, thus corresponding to the classical electron radius. Later, H.S.\,Snyder \cite{Snyder} in 1947 introduced
a minimum length by ``quantizing'' the spacetime, thus anticipating of more than half a century the so-called ``non-commutative geometry'' models (see below). Spacetime coordinates are represented by quantum operators and, in Snyder's approach, these operators have a discrete spectrum, thus entailing a discrete interpretation of spacetime. Subsequently, also T.D.\,Lee \cite{Lee} introduced an effective time discretization on the basis of the necessarily finite number of measurements performable in any finite interval of time.

\

\noindent The concept of non-continuous spacetime has recently returned into fashion in GUT's \cite{GUTs}, in string theories \cite{String}, in quantum gravity \cite{QG} and in the approaches regarding spacetime either as a sort of quantum ether or as a spacetime foam \cite{Foam}, or even as endowed with a non-commutative geometry (like in Deformed or Double Special Relativity \cite{NCG}\footnote{It is interesting to point out that in the first of Refs.\cite{NCG} it is stated that ``the special role of the time coordinate in the structure of $k$-Minkowski spacetime forces one to introduce an element
of discretization in the time direction''.}). In particular, M-Theory, Loop Quantum Gravity \cite{Alfaro,LQG} and Causal Dynamical Triangulation \cite{Loll} lead to postulate an essentially discrete and \textit{quantum} spacetime, where (as expected from the uncertainty relations) fundamental momentum and mass-energy scales naturally arise, in addition to $\hbar$ and $c$.
In contemporary physics the most important case of an elementary space scale is doubtless the ``Planck length'' $\dis \ell_{\rm P}=\sqrt{\frac{\hbar G}{c^{3}}} \simeq 1.62\cdot 10^{-33}$\,cm; for example, it plays a fundamental role in string theory, where it is defined as the minimum length of a typical string. As a consequence, any space distance smaller than $\ell_{\rm P}$ is deprived of any physical meaning or possibility of being measured. M.\,Planck himself, early in 1899 \cite{Planck} proposed a system of units to be used in fundamental physics (a ``System of Natural Units''), where fundamental units are the Planck length, the Planck time, and the Planck mass. Although Planck did not know yet quantum mechanics and general relativity, such scales mark out the regions where these
theories become, in a sense, reciprocally in contrast, so that a proper quantum gravity theory is required. At the present the Planck length is the minimum spacetime scale present in physics, at which
we expect a substantial ``unification'' of high energy microphysics and early cosmology.

\

\noindent This is, in short terms and until now, the known story of the introduction in physics of ``fundamental'' constants with the meaning of elementary lengths or times. However, with the thorough study, started in recent years, of the copious set of unpublished notes left by another protagonist of theoretical physics \cite{EMReview}, Ettore Majorana \cite{EMBio}, we have been acquainted with some unknown contributions pointing out his own elaboration about the arguments just described above. In the remaining part of this paper, we will then describe and comment such contributions, present in the so-called {\it Volumetti} \cite{Volumetti} and {\it Quaderni} \cite{Quaderni}, containing study and research notes which were not published by Majorana.

In these booklets, among many different topics the author studied extensively several arguments of electrodynamics (both at a classical and at a quantum level). Many of these regarded topics and methods ``usually'' discussed in the scientific workplaces of 1930s, although they were dealt with in a very original manner and, sometimes, with extraordinary results (see, for example, Ref. \cite{EMelectro}). Here, however, we will focus just on three of these notes, which result to be particularly relevant for the topics considered here. A characteristic of the Majorana notes is that they always concern specific examples rather than generic theoretical issues. In the following section, then, we report the study performed by Majorana about the scattering of particles by a {\it quasi}-Coulombian potential, where a non vanishing radius for the scatterer is considered. In Sect.\,3 we instead discuss the introduction of an intrinsic time delay in the propagation of the (classical) electromagnetic field, considered by Majorana in order to take into account the possible effect on the electromagnetic potentials of a fundamental constant length. Finally, in Sect.\,4 we will describe an interesting (though leading to incorrect results) attempt to find a relation between the fundamental physical constants $e$, $\hbar$ and $c$. Our conclusions and outlook then follow.

\section{Quasi-Coulombian scattering}

\noindent The set of Majorana's \textit{Quaderni} opens with the study of the problem of the scattering of particles from a quasi-Coulombian potential of the form
\begin{equation}
\displaystyle V(r)=\frac{k}{\sqrt{r^{2}+a^{2}}} \, ,
\label{quasi}
\end{equation}
where $k$ is a positive constant (related to the electric charge of the potential source) and $a$ is, according to Majorana, the ``magnitude of the radius of the scatterer''. We do not know precisely the motivations for such a study (likely, for applications to atomic and nuclear physics problems; see below), but, as a matter of fact, the original intention of the author was to extend the well-known Rutherford formula for the scattering of a beam of particles (with charge $Z^\prime e$ end mass $m$) from a given body (of charge $Z e$). In this case we have a pure Coulomb scattering, and the cross section (number of scattering particles at an angle $\theta$ per unit time and solid angle) is given by the classical formula
\begin{equation}
\displaystyle f(\theta)=\frac{Z^{2}Z^{\prime 2} e^{4}}{4m^{2}v^{4}\sin^{4}\theta/2}=\frac{Z^{2}Z^{\prime 2} e^{4}}{16 T^{2}\sin^{4}\theta/2} \, ,
\label{ruth}
\end{equation}
where $v$ is the velocity of the incident particles and $T$ is their kinetic energy. Majorana deduced the above Rutherford formula in his \textit{Volumetti}, using both classical mechanics arguments and the quantum Born approximation method \cite{Volumetti}.
As well-known, Eq.\,(\ref{ruth}) was employed for the first time by E. Rutherford in his experiments on $\alpha$ particles impinging on a gold target, aimed at explaining the structure of the atom (with or without a compact nucleus at its center).

The modification of the Coulomb scattering potential, considered by Majorana in Eq.\,(\ref{quasi}), certainly introduces an improvement in the description of the physical phenomenon, with the introduction of the dimensions of the scattering center (in the Rutherford formula assumed to be pointlike, $a=0$), but the story does not end here. In fact, the most simple approximation in this line of thinking is to consider the scattering center as a uniformly charged sphere of radius $a$ but, in such a case, the potential outside the sphere is strictly Coulombian (thanks to the Gauss law), and Eq.\,(\ref{quasi}) would not apply. Of course, at the time when Majorana performed his calculations, it was known that the nucleus of a given atom is not uniformly charged, being formed by individual particles, but, again, such a situation is not described by the simple formula in Eq.\,(\ref{quasi}). An example is the Yukawa potential \cite{Yukawa}, where the Coulomb potential acquires a screening factor with an exponential form ruled by one more parameter, giving the range of nuclear forces. In any case, it is striking that the only application of Eq.\,(\ref{quasi}) reported in the \textit{Quaderni} \cite{Quaderni} was to the hydrogen atom, where just one proton forms its nucleus, thus being considered effectively as a uniformly charged particle. Thus the reasoning of Majorana behind Eq.\,(\ref{quasi}) has a different starting point, upon which we will briefly speculate below. For the moment, we only note that, from a strictly mathematical point of view, the introduction of a non-vanishing radius for the scattering center has the effect of regularizing the Coulomb potential which, otherwise, would diverge for $r\rightarrow 0$ (however, by contrast to other cases, Majorana did not restore the full Coulomb potential by taking the limit $a\rightarrow 0$ at the end of his calculations, but always maintained in the present case a finite value for $a$).

In the problem proposed, Majorana studied the deviations from pure Coulomb scattering by parameterizing it with the ratio $i/i_{R}$ of the effective scattering intensity under an angle $\theta$ (i.e. the flux of scattered particles per unit surface (normal to the incident direction) and per unit time) with respect to that deduced in the Rutherford approximation. Besides the radius $a$, this ratio depends also on the energy and momentum of the incident particles, but Majorana preferred to parameterize these by means of the scattering parameter (or, as denoted by himself, the ``minimum approach distance'') $b$ in the Coulomb limit, defined by $K/b=T$, and from the wavelength $\lambda$ of the free particle. By setting $\displaystyle \alpha=a/(\lambda/2 \rm{\pi})$, $\displaystyle \beta=b/(\lambda/2 \rm{\pi})$, he obtained:
\begin{equation}
\displaystyle i=f(\alpha,\beta,\theta) \, i_{\rm{R}} \, .
\label{factor}
\end{equation}
Of course, in the pure Coulomb limit, the Rutherford formula should be recovered, so that $f(0,\beta,\theta)=1$. The calculations then proceed at evaluating the function $f(\alpha,\beta,\theta)$ by keeping $\alpha$ fixed (that is, for fixed scatterer size) and considering the limit $\beta\rightarrow 0$ (for scattering particles with increasing momentum, approaching nearer and nearer the center). The wavefunction of the system is, then, evaluated perturbatively  by expanding it in a series ruled by the experimental parameter $\beta$, at zeroth order ($\beta=0$) being used the WKB method \cite{WKB}. After some passages\footnote{In order to avoid convergence problems in the usage of the Green method, Majorana assumes that the scattering force acts only for distances less then a quantity $R$, and than let $R\rightarrow\infty$ at the end of calculations. The potential in Eq.\,(\ref{quasi}) is thus replaced during calculations, by $\displaystyle k \left(\frac{1}{\sqrt{r^{2}+a^{2}}}-\frac{1}{R}\right)$}.
He obtains the following analytic result in the approximation considered $(\rho=r/(\lambda/2\pi))$
\begin{equation}
\displaystyle f(\alpha,\beta,\theta)=2\sin \frac{\theta}{2}\int_{0}^{\infty}\frac{\rho}{\sqrt{\rho^{2}+\alpha^{2}}}\sin \left[2\rho\sin\frac{\theta}{2}\right]\rm{d}\rho \, ,
\end{equation}
which can be expressed in term of the Bessel $k$ function  $\displaystyle k\left(\alpha \sin {\theta}/{2}\right)$ \cite{AS}. In this approximation Majorana then finds that the actual scattering intensity may be substantially different from that of the Rutherford formula, for backward scattering and provided that the radius $a$ is appreciably different from zero (the ratio $i/i_{\rm{R}}$ approaching zero for $\theta=\rm{\pi}$ and increasing $\alpha$).

This remarkable result is not explicitly reported in the \textit{Quaderni}, but Majorana gives an attempt to tabulate the radial wavefunction $u_{\ell}$, satisfying the equation
\begin{equation}
\displaystyle u^{\prime}_{\ell}+\left(1-\frac{\beta}{\sqrt{\rho^{2}+\alpha^{2}}} - \frac{\ell(\ell +1)}{r^{2}}\right)u_{\rm{e}}=0 \, ,
\end{equation}
obtained numerically with the method of the particular solutions. Although numerical solutions are searched only for the particular case of $\ell=0$ and $\beta=0.4$ (and not $\beta=0$), here the interesting point is that Majorana explicitly reports that ``for the hydrogen atom we consider the values $\beta=0.4,0.5,0.6,0.7$ and $\alpha=0,0.2,0.4,0.6,0.8,1$''. Unfortunately, no further discussion is given in the {\it Quaderni}; the key point is, however, the fact that Majorana uses $\lambda/2\rm{\pi}$ as the length scale for \textit{both} the lengths $a$ and $b$. Now, it is very natural to measure the scattering parameter (or ``the minimum approach distance'') in terms of the free particle wavelength or, equivalently, (the inverse of) the free particle momentum $(\lambda=h/p)$, since it is obvious that increasing the momentum (or decreasing the probe wavelength), the incident particle approaches nearer and nearer the scattering center, thus decreasing $b$. The same would not apply, instead, to the radius of the scattering center which should be independent of the incident particle properties, \textit{unless} Majorana considers \textit{effective}, momentum-dependent, dimensions for the scattering center.
In such a case, the attention is evidently shifted from the investigation of an actual, particular physical system considered (particles in a scattering potential) to the study of more general properties of the background field or space. We will come back later on this point.

Finally, we conclude this section by mentioning also another modification of the Coulomb potential considered by Majorana \cite{Quaderni}, reminiscent of the Gamow potential for the description of $\alpha$-decay process, namely
\begin{eqnarray}
&& V = \left\{
\begin{array}{ll}
V_{0}, & \quad \textrm{for} \ r <R, \\
\ \\
\dis\frac{k}{r} \, , & \quad \textrm{for} \ r>R . \\
\end{array}
\right.
\label{bound}
\end{eqnarray}
Here the role of the scattering center radius is played by the range $R$, $\alpha=R/(\lambda/2\pi)$, and one more parameter enters, that is the depth $V_{0}$ of the potential, measured with respect to the kinetic energy of the incident particles, $\displaystyle A=V_{0}/T$.\\
Although, in this case, the calculations are only sketched, the key points envisaged above are as well present. Now the correction function in Eq.\,(\ref{factor}) is replaced by
\begin{equation}
\displaystyle \frac{i}{i_{\rm{R}}}=f\left(\frac{V_{0}}{T},\frac{R}{\lambda/2\pi},\frac{b}{\lambda/2\pi},\theta\right)
\end{equation}
where ``for the hydrogen'' Majorana considers the same values as before for the $\alpha,\beta$ parameters (except, obviously, $\alpha=0,0.2$), while $A=$2,1.5,1,0.5,0,-0.5,-1,-1.5,-2,-2.5,-3,...,-8. Interestingly, besides negative values for $V_{0}$ (as in the Gamow model), Majorana also considers few positive values for the depth of the potential. As above, however, the very notable point is that both the potential parameters $V_{0}$ and $R$ appear to be strictly related to the energy and momentum of the probe particles.

For both the cases studied, in the presence of continuum states (Eq.\,(\ref{quasi})) or bound states (Eq.\,(\ref{bound})), the reasoning is thus the same. This is even more intriguing if compared to the comprehensive study of the scattering from a pure Coulomb potential, performed by Majorana \cite{Quaderni} some pages after those discussed here up to now. Indeed, in this case the length unit is no more the free particle wavelength, the Bohr radius being introduced (and the energy unit is the Rydberg), evidently denoting possible applications to atomic problems. From what discussed above, it is thus clear that Majorana's original motivations for the studies of quasi-Coulombian scattering are different from those standard related to atomic and nuclear physics.

\section{Retarded electromagnetic fields and the introduction of a fundamental length}

\noindent Some light on this issue may come from other pages of Majorana's \textit{Quaderni} \cite{Quaderni}, where the possibility is considered of introducing an intrinsic constant time delay (or, equivalently, an intrinsic space constant) in the expression for the  retarded electromagnetic fields. Here the treatment is fully classical, and goes as follows.

The starting point is the wave equation satisfied by any component of the electromagnetic potentials, denoted generically with $f(x,y,z,t)$, and then the evaluation of the D'Alembert operator for the standard retarded field denoted by
\begin{equation}
\displaystyle {\varphi(x,y,x,t)=f\left(x,y,z,t-\frac{r}{c}\right)\equiv\overline{f(x,y,z,t)}}
\label{phi} \, .
\end{equation}
The known result is, of course, the following:
\begin{equation}
\displaystyle \overline{\square f}=\nabla^{2}\varphi+\frac{2}{rc}\dot{\varphi}+\frac{2}{c}\frac{\partial^{2}\varphi}{\partial r \partial t} \, ,
\label{Dalembert}
\end{equation}
where a dot indicates time differentiation. Majorana then introduces an intrinsic space constant $\varepsilon$, so that Eq.\,(\ref{phi}) is replaced by
\begin{equation}
\displaystyle \varphi(x,y,z,t)=f\left(x,y,z,t-\frac{\sqrt{r^{2}+\varepsilon^{2}}}{c}\right) = \tilde{f}(x,y,z,t) \, .
\end{equation}
The D'Alembertian term entering into the wave equation is thus
\begin{equation}
\displaystyle \widetilde{\square f}=\nabla^{2}\varphi-\frac{\varepsilon^{2}}{c^{2}(r^{2}+\varepsilon^{2})}\ddot{\varphi}+\frac{2r^{2}+3\varepsilon^{2}}{c(r^{2}+\varepsilon^{2})^{3/2}}\dot{\varphi}+\frac{2r}{c\sqrt{r^{2}+\varepsilon^{2}}}\frac{\partial^{2}\varphi}{\partial r \partial t} \, ,
\end{equation}
which replaces Eq.\,(\ref{Dalembert}). Majorana also introduces explicitly a time delay $\tau$ which, in his view, is a ``universal constant'' taking the value $\tau=0$ classically; in his calculations, however, he always uses the length
\begin{equation}
\displaystyle \varepsilon=\tau c
\end{equation}
(already introduced), so that we here follow this line of reasoning.

The wave equation is apparently solved by using the Green method; denoting with $S^{\prime}$ the generic source function (charge density $\rho$ or current density ${\Jbf}$), the modified generic potential $\Phi$, solution of the wave equation, assumes now the following from:
\begin{equation}
\displaystyle \Phi=\int\frac{1}{\sqrt{R^{2}+\varepsilon^{2}}} \, S\left(x^{\prime},y^{\prime},z^{\prime},t-\frac{\sqrt{R^{2}+\varepsilon^{2}}}{c}\right)\rd x^{\prime}\rd y^{\prime}\rd z^{\prime}
\label{potential}
\end{equation}
(with $R=\left|\rbf-\rbf^{\prime}\right|$, as usual). Majorana then ends his explicit calculations with the writing of Eq.\,(\ref{potential}) expanded up to second order in $\varepsilon$ for $\varepsilon \rightarrow 0$ (thus approaching the classical limit, as of course coming from experiments):
\bea
\Phi &=& \int\frac{1}{R}S\left(x^{\prime},y^{\prime},z^{\prime},t-\frac{R}{c}\right){{\rm d}x}^{\prime}{{\rm d}y}^{\prime}{{\rm d}z}^{\prime} \nonumber
\\
& & - \, \varepsilon^{2}\left\{\int\frac{1}{2R^{3}}S\left(x^{\prime},y^{\prime},z^{\prime},t-\frac{R}{c}\right){{\rm d}x}^{\prime}{{\rm d}y}^{\prime}{{\rm d}z}^{\prime} \right. \nonumber \\
& & \left. + \int\frac{1}{2R^{2}c}S^{\prime}\left(x^{\prime},y^{\prime},z^{\prime},t-\frac{R}{c}\right){{\rm d}x}^{\prime}{{\rm d}y}^{\prime}{{\rm d}z}^{\prime}\right\}+\ldots \, .
\eea
Some comments are now in order. Unfortunately, again we do not know the motivations for such a study, but it is intriguing  the following observation. By taking the ordinary electric monopole limit \cite{Jackson} into Eq.\,(\ref{potential}), with $\displaystyle S(\rbf,t-\frac{r}{c})=\rho(\rbf)$, $\left|\rbf-\rbf^\prime\right|\cong \left|\rbf\right|=r$ and $\displaystyle \int \rho(\rbf^\prime)\rd x^\prime\rd y^\prime\rd z^\prime=Q$ being the electric charge, one can easily obtain
\begin{equation}
\displaystyle \Phi=\frac{Q}{\sqrt{r^{2}+\varepsilon^{2}}} \, ,
\label{Phi}
\end{equation}
that is exactly the potential considered in Eq.\,(\ref{quasi}).

Although Eq.\,(\ref{Phi}) is not explicitly reported in the \textit{Quaderni}, it seems very likely that the topics discussed here and in the previous section (and considered in no other place in the \textit{Quaderni}) are related. If so, the ``magnitude of the radius of the scatterer'' in the scattering problem considered above, whose interpretation in terms of the physical dimensions of the scattering center led to difficulties, should be interpreted instead as the ``universal'' space constant introduced here. In this case Majorana's original intention for the study of quasi-Coulombian scattering was not that of exploring the properties of the scattering particle (an atom or its nucleus or even other elementary particles) but rather that of investigating the underlying properties of the physical space. The intrinsic space constant then plays the role of a fundamental length, like the one introduced by Planck in 1899 or that conjectured by Heisenberg in 1938 (see the the Introduction). We do not know if Majorana effectively thought to a given particular value but, in any case, to the best of our knowledge, with Majorana it is the first time that such a detailed study was effectively undertaken, as early as in the 1930s.

\section{A relation among the fundamental constants}

\noindent The great interest of Majorana for electrodynamics \cite{EMelectro}, in its classical or (especially) quantum form, is very well documented in his writings (see mainly Refs.\,\cite{Volumetti} and \cite{Quaderni}), so that even the particular studies described above do not seem at all surprising.  Nevertheless, the line of thinking of such studies goes well beyond mere applications or generalizations of standard electrodynamics, as shown in the previous pages, involving questions about fundamental constants. It is intriguing that yet another time Majorana reasoned about such fundamental questions, though in an (apparently) different context, when he was still a student, as early as in 1928.
Indeed, in a page of his \textit{Volumetti} (see section 31 of Volumetto II, in Ref.\,\cite{Volumetti}), Majorana attempted to find a relation between the fundamental constants $e$, $h$ and $c$, or, more explicitly, an expression for the elementary charge $e$ appearing in the Coulomb law, describing, for example, the electrostatic force between two electrons. The reasoning of Majorana goes as follows. In the region of space surrounding the two electrons, which are at a distance $\ell$ apart, an electromagnetic field is present that, ``in some sense'' (according to Majorana's wording), is quantized. Here we should point out the first interesting thing since, in what follows (see below), it is evident that what is considered ``quantized'' by Majorana is undoubtedly the electromagnetic field acting among the two electrons.
Nevertheless, differently from what appearing elsewhere in his notes, Majorana does not refer explicitly to an electromagnetic \textit{field}, but rather to the 'ether' surrounding the two electrons, that is (apparently) the physical space itself (in the improper language of that time). The two electrons are assumed to interact among them by means of a point particle (what we could term the quantum of the electromagnetic field) moving with a group velocity equal to the speed of light $c$. Such particle is further assumed to propagate freely from one electron to the other, whereas it inverts its motion when ``colliding'' with the electrons.
The momentum of the quantum particle is deduced by a sort of Sommerfeld quantization rule
$$
\left|p\right|=\frac{nh}{2\ell} \, ,
$$
with $n=1$. The ``kick'' received by the electron is therefore $\dis 2\left|p\right|=h/\ell$, and since the number of collisions per unit time is $\dis 1/T=c/2\ell$, the magnitude of the force acting on each electron is then $\dis F=2\left|p\right|/T=hc/2\ell^{2}$.
By equating this expression to the Coulomb law, Majorana finally deduces the expression he searched for the elementary charge $\dis e=\sqrt{hc/2}$ which, however, as pointed out explicitly by himself, gives a value ``21 times greater than the real one''. Though leading to results far from the truth, what discussed by Majorana is interesting for many reasons, one of which already mentioned above. Indeed, first of all, the mechanical model considered here is the first one to have been proposed as an attempt to deduce the value of the electric charge (the famous paper by Dirac in Ref.\,\cite{Dirac} was published few years after). However, as it should be evident from what reported above, the interpretation given by Majorana of the \textit{quantized} electromagnetic field substantially coincides with that introduced more than a decade later by R.P. Feynman in quantum electrodynamics \cite{Feynman}, the point particles exchanged by electrons being assumed to be the photons. It is then remarkable that fundamental questions regarding the constants of Nature discussed properly only in more recent times, were addressed in pioneering works by Majorana as early as at the end of 1920s, where, on the other hand, the key ideas of quantum field theory were already present.

\section{Conclusions}

\noindent The problem of the existence and of the effects of a fundamental length dates back to the early years of the last century. As reviewed in the introduction, this concept was introduced and rediscovered in physics several times since then, in order to solve difficulties of different kind as, e.g., in regularizing some divergent results. It is, however, quite notable the presence of ``side" effects associated to (or, rather, postulated about) the existence of such elementary space intervals, which could allow, in Heisenberg's view for example, the derivation of the mass of the particles, or even play a key role in the physical interpretation of the spin variable.

Strictly related to the introduction of a fundamental length is, on the other hand, the assumption of the existence of a fundamental time, although such hypothesis is less fashioned among physicists than the first one.

In more recent times, the present subject is embedded into the more general framework of the quantization of the spacetime, where attempts to provide a well-defined quantum theory of gravity manifest into different theoretical approaches, with the common denominator of a fundamental length scale.

In this paper we have shown that the known picture about this intriguing concept has been further enriched by several, yet unknown, contributions by Ettore Majorana. As early as the beginning of 1930s, indeed, the Italian physicist not only conjectured (as already Planck and, later, others did) the existence of both an elementary length and of a fundamental time scale, but also investigated in some detail the immediate physical consequences of such hypothesis. In this respect, his contributions once again \cite{EMReview} reveal a farsighted intuition, and result to be quite precious pieces of research both on the historical point of view and for present day theoretical investigations.

Majorana considered, for example, the scattering from electrically charged particles when the underlying properties of the space are such that an intrinsic length exists, resulting in the modification of the pure Coulomb potential as in Eq.\,(\ref{quasi}) or (\ref{Phi}). Even just at the classical level, he obtained the same effect by introducing an intrinsic time delay, which is considered a ``universal constant", in the expression for the retarded electromagnetic fields.

At the quantum level, instead, it is quite interesting still another contribution by Majorana about fundamental constants, aimed at finding a relation between $e$, $\hbar$, $c$ or, rather, aimed at deducing the value of the elementary charge. Such a scope was pursued, few years later, in a famous paper by Dirac \cite{Dirac}, although based on a different theoretical scheme and with different results. Indeed, the particular {\it numerical} prediction by Majorana (as opposed to the particular {\it theoretical} prediction of the existence of magnetic monopoles by Dirac) is far from the truth, but the key idea behind it anticipated of more than a decade the assumption, introduced by Feynman in quantum electrodynamics, of photon exchange during electromagnetic processes.

We expect that Majorana's important ideas and applications recovered from the oblivion through this paper will bring some benefit also to the present day abstract theoretical research on the fundamental properties of the physical spacetime.

\vspace{0.5cm}

\noindent {\large{\bf Acknowledgments}}

\vspace*{0.2cm}

\noindent The authors warmly thank E.\,Recami for interesting discussions and suggestions, while  acknowledge the very kind collaboration by M.G.\,Cammarota and M.\,Longo.

\

\end{document}